\documentclass[conference,a4paper]{IEEEtran}
\IEEEoverridecommandlockouts
\usepackage{cite}
\usepackage{amsmath,amssymb,amsfonts}
\usepackage{algorithmic}
\usepackage{graphicx}
\usepackage{textcomp}
\usepackage{xcolor}
\def\BibTeX{{\rm B\kern-.05em{\sc i\kern-.025em b}\kern-.08em
    T\kern-.1667em\lower.7ex\hbox{E}\kern-.125emX}}

\usepackage{afterpage}

\usepackage{tikz, calc}
\usepackage{circuitikz}
\usepackage{subfig}
\usetikzlibrary {arrows.meta, decorations.pathmorphing}
\usetikzlibrary{positioning, arrows, calc}
\usepackage{listofitems}

\begin{document}

\title{\textit{Transporter}: A 128$\times$4 SPAD Imager with On-chip Encoder for Spiking Neural Network-based Processing\\}

\author{\IEEEauthorblockN{ Yang Lin \qquad Claudio Bruschini \qquad Edoardo Charbon}
\IEEEauthorblockA{\textit{Advanced Quantum Architecture Lab} \\
\textit{École Polytechnique Fédérale de Lausanne}\\
Neuchâtel, Switzerland \\
\texttt{\{yang.lin, claudio.bruschini, edoardo.charbon\}@epfl.ch}}
}

\maketitle

\begin{abstract}
Single-photon avalanche diodes (SPADs) are widely used today in time-resolved imaging applications, however traditional architectures rely on time-to-digital converters (TDCs) and histogram-based processing, leading to significant data transfer and processing challenges. Previous work based on recurrent neural networks has realized histogram-free processing. To further address these limitations, we propose a novel paradigm that eliminates TDCs by integrating in-sensor spike encoders. This approach enables preprocessing of photon arrival events in the sensor while significantly compressing data, reducing complexity, and maintaining real-time edge processing capabilities. A dedicated spike encoder folds multiple laser repetition periods, transforming phase-based spike trains into density-based spike trains optimized for spiking neural network processing and training via backpropagation through time. As a proof of concept, we introduce \textit{Transporter}, a 128$\times$4 SPAD sensor with a per-pixel D flip-flop ring-based spike encoder, designed for intelligent active time-resolved imaging. This work demonstrates a path toward more efficient, neuromorphic SPAD imaging systems with reduced data overhead and enhanced real-time processing.
\end{abstract}

% \begin{IEEEkeywords}
% component, formatting, style, styling, insert
% \end{IEEEkeywords}

\section{Introduction}

\tikzstyle{rect} = [rectangle, rounded corners, minimum width=2cm, minimum height=1cm,text centered, draw=black]

\tikzset{
    position/.style args={#1:#2 from #3}{
        at=(#3.#1), anchor=#1+180, shift=(#1:#2)
    }
}

\definecolor{oceanboatblue}{rgb}{0.0, 0.47, 0.75}
\definecolor{napiergreen}{rgb}{0.16, 0.5, 0.0}
\definecolor{neoncarrot}{rgb}{1.0, 0.64, 0.26}
\definecolor{candypink}{rgb}{0.89, 0.44, 0.48}

\tikzstyle{mynode}=[thick,draw=blue,fill=blue!20,circle,minimum size=10]

\tikzset{>=latex} % for LaTeX arrow head
\colorlet{myred}{red!80!black}
\colorlet{myblue}{blue!80!black}
\colorlet{mygreen}{green!60!black}
\colorlet{mydarkblue}{blue!40!black}
\tikzstyle{node}=[thick,circle,draw=myblue,minimum size=22,inner sep=0.5,outer sep=0.6]
\tikzstyle{node in}=[node,green!20!black,draw=mygreen!30!black,fill=mygreen!25]
\tikzstyle{node hidden}=[node,blue!20!black,draw=myblue!30!black,fill=myblue!20]
\tikzstyle{node out}=[node,red!20!black,draw=myred!30!black,fill=myred!20]
\tikzstyle{connect}=[thick,mydarkblue] %,line cap=round
\tikzset{ % node styles, numbered for easy mapping with \nstyle
  node 1/.style={node in},
  node 2/.style={node hidden},
  node 3/.style={node out},
}
\def\nstyle{int(\lay<\Nnodlen?min(2,\lay):3)} % map layer number onto 1, 2, or 3

\definecolor{skyblue}{rgb}{0.53, 0.81, 0.92}

\begin{figure*}
    \centering
    \begin{circuitikz}[x=1cm,y=1cm,scale=0.9, transform shape]
    % light source

    \node [outer sep=0pt,circle, fill=neoncarrot, minimum width=0.3cm, minimum height=0.3cm] (laser) at (0, 0) {};

    \foreach \i in {0,1,...,11}{
    \draw [very thick,neoncarrot] (laser) -- ++(30*\i:0.55cm);
    \draw [very thick,neoncarrot] (laser) -- ++(15+30*\i:0.35cm);
    }

    % sample
    \node (samplecenter) at (-3,1) {};
    % \node [position=165:{3cm} from laser,] (samplecenter) {};
    \node [circle, fill=green, position=-30:{0cm} from samplecenter] {};
    \node [circle, fill=blue, position=90:{0cm} from samplecenter] {};
    \node [circle, fill=red, position=210:{0cm} from samplecenter] {};

    % piccolo
    \node [rect,draw, minimum width=1.2cm,minimum height=1.2cm,above=2cm of laser.center,label=below:SPAD] (spad) {};
    \draw (spad.south) to [Do, fill=skyblue] (spad.north);

    \node [rect, draw, text width=2.2cm] (tdc) at (3.5, 3.2) {Time-to-digital Converter};
    \node [rect, draw, right=1cm of tdc] (histogram) {Histogram};
    \node [rect, draw, right=1cm of histogram] (proc) {Data Processing};

    \node [rect, draw, text width=2cm] (encoder) at (3.5, -0.2) {Spike Encoder};

    \node [rect, draw, text width=2.2cm] (tdc2) at (3.5, 1.7) {Time-to-digital Converter};
    \node [rect, draw, right=1cm of tdc2, text width=2.5cm] (rnn) {Recurrent Neural Networks};

    % spiking neural network
    \readlist\Nnod{1,4,4,4,1}
    \foreachitem \N \in \Nnod{ % loop over layers
    \edef\lay{\Ncnt} % alias of index of current layer
    \message{\lay,}
    \pgfmathsetmacro\prev{int(\Ncnt-1)} % number of previous layer
    \foreach \i [evaluate={\y=\N/2-\i; \x=\lay; \n=\nstyle;}] in {1,...,\N}{ % loop over nodes
      
      % NODES
      \node[node \n,minimum size=8] (N\lay-\i) at (5.5+0.5*\x,0.4*\y) {};
      %\node[circle,inner sep=2] (N\lay-\i') at (\x-0.15,\y) {}; % shifted node
      %\draw[node] (N\lay-\i) circle (\R);
      
      % CONNECTIONS
      \ifnum\lay>1 % connect to previous layer
        \foreach \j in {1,...,\Nnod[\prev]}{ % loop over nodes in previous layer
          \draw[connect] (N\prev-\j) -- (N\lay-\i); % connect arrows directly
          %\draw[connect arrow] (N\prev-\j) -- (N\lay-\i'); % connect arrows to shifted node
        }
      \fi % else: nothing to connect first layer
      
    }
    }

    \node [rect, draw, minimum width=2.7cm, minimum height=1.8cm, label=below:SNN] (snnnode) at (7, -0.2) {};

    \node [circle, minimum width=1.5cm, label=below:Laser] (laserph) at (laser.center) {};
    \node [circle, minimum width=1.1cm, label=left:Object] (sampleph) at (samplecenter.center) {};

    \draw [->, decorate, decoration={snake}] (laserph) -- node[below,sloped] {illumination} (sampleph);
    \draw [->, decorate, decoration={snake}] (sampleph) -- node[below,sloped] {emission} node[above,sloped] {reflection or} (spad);

    \draw [->, rounded corners=5pt, thick, candypink] (spad.east) -- ++(0.5,0) |- (tdc.175);
    \draw [->, rounded corners=5pt, thick, candypink] (spad.east) -- ++(0.5,0) |- (tdc2.175);
    \draw [->, rounded corners=5pt, thick, candypink] (spad.east) -- ++(0.5,0) |- (encoder.175);

    \draw [->, rounded corners=5pt, thick, neoncarrot] (laserph.east) -- ++(0.9,0) |- (tdc.185);
    \draw [->, rounded corners=5pt, thick, neoncarrot] (laserph.east) -- ++(0.9,0) |- (tdc2.185);
    \draw [->, rounded corners=5pt, thick, neoncarrot] (laserph.east) -- ++(0.9,0) |- (encoder.185);

    \draw [->, ultra thick] (tdc) -- node[above, text width=1.5cm, scale=0.8, align=center]{Large volume multibit data} (histogram);
    \draw [->, ultra thick] (tdc2) -- node[above, text width=1.5cm, scale=0.8, align=center]{Large volume multibit data} (rnn);
    \draw [->, very thick] (histogram) -- node[above, text width=1.5cm, scale=0.8, align=center]{Multibit data} (proc);
    \draw [->, thick] (encoder) -- node[above, text width=1cm, scale=0.8, align=center]{1-bit data} (snnnode);

    % image
    \node[rectangle, rounded corners, minimum width=1.5cm, minimum height=1.5cm, right=1cm of snnnode, fill=lightgray, text=white] (img) {Image};

    \draw [thick, ->] (snnnode) -- (img);
    \draw [thick, ->, rounded corners=5pt] (rnn) -| (img);
    \draw [thick, ->, rounded corners=5pt] (proc.east) -- ++ (0.5, 0) |- (img);
    
    \end{circuitikz}
    \caption{Workflow of the traditional SPAD TCSPC system, RNN-based system, and the proposed \textit{Transporter} system. In a traditional SPAD TCSPC system, the laser illuminates the object and sends a reference signal to the TDCs. The SPAD detects the reflected or emitted light from the object. The reference and detection signals from the SPAD are directed to TDCs for time-tagging. The timestamps are often histogrammed and transmitted to a PC/Mac for data processing, where the image is reconstructed. For the RNN-based system, the timestamps are directly processed within or near the sensor, then the images are sent to PC/Mac. In the proposed \textit{Transporter} imager, the on-chip spike encoder synchronizes with the laser, and receives pulses upon photon arrival. The encoder keeps collecting photons until the readout is launched, then the pulses are sent to the SNN, which is deployed either on the chip or the FPGA, where the images are reconstructed.}
    \label{fig:workflow}
\end{figure*}
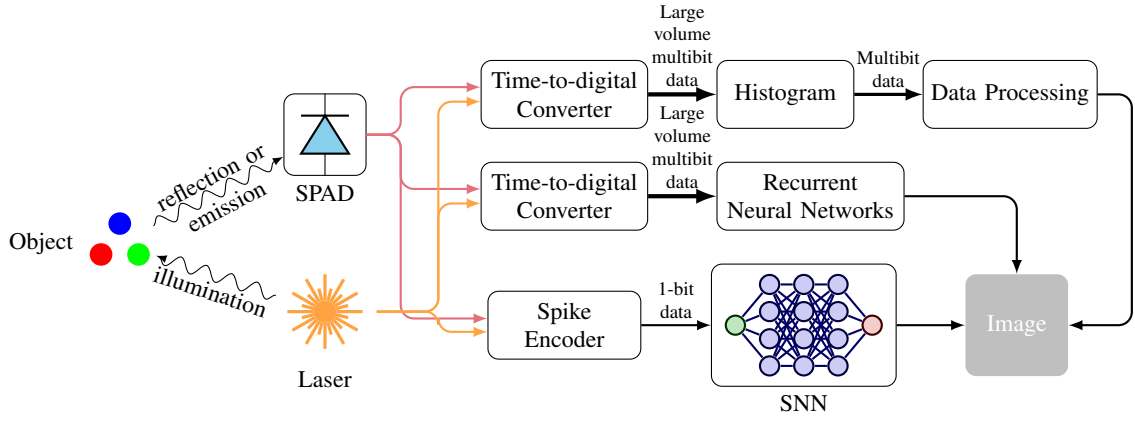

Single-photon avalanche diodes (SPADs) are highly sensitive photodetectors capable of detecting individual photons with precise timing resolution. They have become widely used in time-resolved imaging applications, such as fluorescence lifetime imaging microscopy (FLIM), light detection and ranging (LiDAR), and time-resolved Raman spectroscopy\cite{bruschini2019single}. Traditional SPAD-based time-resolved imaging systems typically employ time-to-digital converters (TDCs) and histogram-building in the workflow \cite{kapusta2015advanced}, as shown in the top path of Fig.~\ref{fig:workflow}. However, the substantial volume of single-photon data generated by these systems presents challenges in data transfer, storage, and processing. To address this problem, artificial neural networks (ANNs) have been proposed as an alternative to histogram-building and/or data processing at the output of the TDC. Recurrent neural networks (RNNs) are deployed on the FPGA to process raw timestamps for LiDAR and FLIM\cite{milanesehistogram, lin2024coupling}. 

In this paper, we go one step further introducing spiking neural networks (SNNs) to eliminate TDCs and to simplify data and image processing. SNNs take spikes as input, making them well suited for processing pulses generated by SPADs without TDC processing, despite training challenges \cite{eshraghian2023training}. Unlike time-tagging methods, only single-bit data are transferred between the SPAD and the SNN, while the data stream can be further compressed using a dedicated spiking encoder. Herein, we propose a novel paradigm that integrates in- or near-sensor spike encoding with SPAD in time-resolved imaging, enabling end-to-end real-time processing at the edge \cite{lin2024spiking}. The spike encoder folds multiple laser repetition periods temporally, converting phase-based spike trains into density-based spike trains, thus significantly compressing the data while retaining the spiking nature essential for SNN processing. The compressed spike trains also facilitate the training based on backpropagation through time (BPTT). Based on the new paradigm, we introduce \textit{Transporter}, an imager inspired by an episode of \textit{Star Trek: The Next Generation} (S6E4, \textit{Relics}). \textit{Transporter} is a proof-of-concept 128$\times$4 SPAD sensor with per-pixel D flip-flop (DFF) ring-based spike encoder, specifically designed for intelligent active time-resolved imaging.

\section{Transporter}

% \subsection{Overview}

\tikzstyle{rect} = [rectangle, rounded corners, text centered, draw=black, anchor=center]

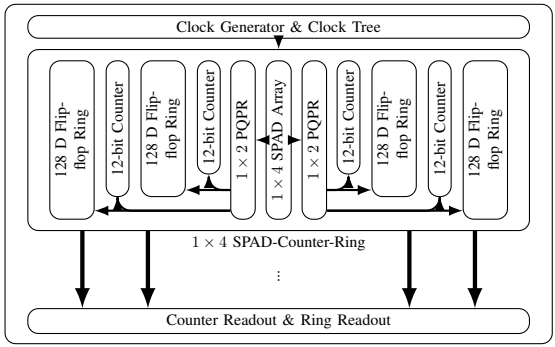
\begin{figure}[!tp]
\centering
\subfloat[High-level schematic of the \textit{Transporter} sensor]{
    \centering
    \begin{circuitikz}[x=1cm,y=1cm,scale = 0.6, transform shape]

    \node[rect, minimum width=12cm, minimum height=7.5cm] (chip) at (0, 2.25) {};
    \node[rect, minimum width=11cm, minimum height=4cm, label=below:{$1\times 4$ SPAD-Counter-Ring}] (row) at (0, 3) {};
    \node[rect, minimum width=11cm, minimum height=0.5cm] (clk) at (0, 5.5) {Clock Generator \& Clock Tree};
    \node[rect, minimum width=3.5cm, minimum height=0.5cm, rotate=90] at (0, 3) (spad) {$1\times 4$ SPAD Array};

    \node[rotate=90] (more) at (0, 0) {...};

    \node[rect, minimum width=11cm, minimum height=0.5cm] (readout) at (0, -1) {Counter Readout \& Ring Readout};

    \draw [->, ultra thick] (row.205) -- (readout.north-|row.205);
    \draw [->, ultra thick] (row.215) -- (readout.north-|row.215);

    \draw [->, ultra thick] (row.335) -- (readout.north-|row.335);
    \draw [->, ultra thick] (row.325) -- (readout.north-|row.325);

    %%%%%%%%
    \node[rect, minimum width=3.5cm, minimum height=0.5cm, rotate=90, right=0.5cm of spad.south, anchor=center] (pqpr) {$1\times 2$ PQPR};
    \node[rect, minimum width=2.5cm, minimum height=0.5cm, rotate=90, right=0.75cm of pqpr.east, anchor=east] (cnt) {12-bit Counter};
    \node[rect, minimum width=3cm, minimum height=0.5cm, rotate=90, right=1cm of cnt.east, anchor=east, minimum width=3cm, text width=2.5cm] (ring) {128 D Flip-flop Ring};
    \node[rect, minimum width=3cm, minimum height=0.5cm, rotate=90, right=1cm of ring.east, anchor=east] (cnt2) {12-bit Counter};
    \node[rect, minimum width=3.5cm, minimum height=0.5cm, rotate=90, right=1cm of cnt2.east, anchor=east, text width=2.5cm] (ring2) {128 D Flip-flop Ring};

    \draw [->, rounded corners=5pt, thick] (pqpr.194) -| (cnt);
    \draw [->, thick] (pqpr.194) -- (pqpr.194-|ring.north);

    \draw [->, rounded corners=5pt, thick] (pqpr.190) -| (cnt2);
    \draw [->, thick] (pqpr.190) -- (pqpr.190-|ring2.north);

    %%%%%%%%
    \node[rect, minimum width=3.5cm, minimum height=0.5cm, rotate=90, left=0.5cm of spad.north, anchor=center] (pqpr2) {$1\times 2$ PQPR};
    \node[rect, minimum width=2.5cm, minimum height=0.5cm, rotate=90, left=0.75cm of pqpr2.east, anchor=east] (cnt3) {12-bit Counter};
    \node[rect, minimum width=3cm, minimum height=0.5cm, rotate=90, left=1cm of cnt3.east, anchor=east, text width=2.5cm] (ring3) {128 D Flip-flop Ring};
    \node[rect, minimum width=3cm, minimum height=0.5cm, rotate=90, left=1cm of ring3.east, anchor=east] (cnt4) {12-bit Counter};
    \node[rect, minimum width=3.5cm, minimum height=0.5cm, rotate=90, left=1cm of cnt4.east, anchor=east, text width=2.5cm] (ring4) {128 D Flip-flop Ring};

    \draw [->, rounded corners=5pt, thick] (spad) -- (pqpr);
    \draw [->, rounded corners=5pt, thick] (spad) -- (pqpr2);

    \draw [->, rounded corners=5pt, thick] (pqpr2.166) -| (cnt3);
    \draw [->, thick] (pqpr2.166) -- (pqpr2.166-|ring3.south);

    \draw [->, rounded corners=5pt, thick] (pqpr2.170) -| (cnt4);
    \draw [->, thick] (pqpr2.170) -- (pqpr2.170-|ring4.south);

    %%%%%%%

    \draw [->, thick] (clk) -- (row);
    % \draw [->, thick] (clk.south|-ring.east) -- (ring.east-|row.north);

    \end{circuitikz}}
    \hfill
% \subfloat[Snapshot of the layout]{
%     \centering
%     \includegraphics[width=0.3\textwidth]{figures/layout_anno.png}}
    \caption{\textit{Transporter} is composed of a clock module, 128 rows of $1\times 4$ SPAD-Counter-Ring, and a readout module.}
    % \caption{\textit{Transporter} is composed of a clock module, 128 rows of $1\times 4$ SPAD-Counter-Ring, and a readout module. The clock generator consists of a ring oscillator and a gating mechanism. The ring oscillator produces a $\sim$1 GHz clock signal, while the gating mechanism ensures that only 128 cycles are supplied to the rings. The clock tree then distributes the clock signal to the four columns of D flip-flop rings. Each row contains four SPADs, with each SPAD paired with a passive quenching and passive recharging circuit, a 12-bit ripple counter, and a D flip-flop ring. The counter and flip-flop ring transmit data to the readout module via four groups of tri-state buffers.}
    \label{fig:high-level}
\end{figure}

\textit{Transporter} is composed of a clock module, 128 rows of 1$\times$4 SPAD-Counter-Ring, and a readout module, as shown in Fig~\ref{fig:high-level}. The clock module, located on the top, provides and distributes the clock to all the rings for operation and readout. For each row of 1$\times$4 SPAD-Counter-Ring, the 1$\times$4 SPADs are located in the middle, while two passive-quenching and passive-recharging (PQPR) circuits and rings are placed on either side. The rings can be organized into four columns, where a tri-state buffer bus goes through each of them and the data are read out by the readout module at the bottom.

% Each SPAD, featuring a 10 $\mu$m circular active area, is paired with a 128-stage D flip-flop ring. The D flip-flops are driven by a gated clock to guarantee 128 cycles per laser repetition period, with a maximum frequency of 1 GHz. Upon photon arrival, the SPAD generates a pulse that is injected into the ring and continues to circulate until it is read out. The ring operates in synchronization with the laser, preserving relative temporal information between photon pulses. The pulses across the sensor are read out sequentially after exposure is complete and sent to spiking processors on the FPGA for inference. Simulation results indicate that the D flip-flop ring-based encoder effectively preserves temporal information and demonstrates high efficiency in SNNs trained for downstream tasks such as FLIM, as shown in Fig.~\ref{fig:result}. 
% The overall design is shown in Figure. Transporter has 128$\times$4 SPADs, each coupled with a passive-quenching and passive-recharging circuit, a 12-bit photon counter, and a 128-stage flip-flop ring. It has an internal ring oscillator-based clock generator, which outputs a clock of 800 MHz. It is also possible to receive an external clock.

\subsection{Clock Module}

The clock module is designed to supply either a high-speed clock to the ring, delivering exactly 128 pulses per laser period, or a low-speed clock for readout. An internal 3-stage NAND-based ring oscillator generates the high-speed clock for ring operation. A separate VDD is supplied to improve the stability of the ring oscillator. Alternatively, an external clock can be selected. The clock is distributed to the four columns of rings with a clock buffer H-tree, and further goes to the 128 rings.

Ensuring exactly 128 pulses per laser period is crucial for maintaining data integrity within the ring. Any deviation, whether an excess or shortage of pulses, would result in the loss of relative timing information between spikes. To address this, a gating mechanism is implemented to enforce the 128-pulse constraint. This mechanism utilizes a 7-bit counter, composed of a 2-bit synchronous counter and a 5-bit asynchronous counter, designed to operate at gigahertz frequencies. Once the counter reaches 128, the clock is disabled and remains so until the next laser synchronization signal arrives, at which point it is re-enabled.

\subsection{SPAD and Pixel Circuits}

The SPAD features $\phi$10 $\mu$m round active area. A PQPR circuit, composed of two NMOS transistors, is employed for each SPAD. The photon detection probability and the dead time can be tuned by adjusting the gate voltage of these transistors. Two inverters are cascaded at the output to digitize the pulse. The pixel pitch is restricted by the height of the \textit{Transporter} ring, which is 25 $\mu$m in our implementation. The horizontal pitch is set equal to the vertical pitch to maintain isotropic pixel spacing, though it can be reduced to 15 $\mu$m. As a result, the fill factor is 12.6\%.

\subsection{Transporter Ring}

The \textit{Transporter} ring functions as the spike encoder as shown in Fig.~\ref{fig:workflow}. In this work, we adopt DFFs as the delay element described in \cite{lin2024spiking}. A simplified schematic is illustrated in Fig.~\ref{fig:transporter}. 128 DFFs are cascaded like a circular shift register, except that an OR gate is inserted between the two ends, with its second input connected to the SPAD output for spike injection. The DFFs are arranged into two rows, with clock buffer H-tree placed on the top and bottom. As mentioned earlier, the accurate clock timing for the DFFs is crucial to the integrity of the spikes. Considering the setup and hold time at four corners, the maximum operation frequency is 1GHz.

In this proof-of-concept \textit{Transporter}, only 128 DFFs are used. As these DFFs store single-bit data, the ring could be saturated in case of high photon flux. Therefore, a ring stopper, based on the ripple counter and the clock gate, is designed to prevent the ring from saturation. With the stopper, the injection of the spike into the ring will be stopped when the photon count rises above 128 or 256.

% \begin{figure}
%     \centering
%     \includegraphics[width=0.6\linewidth]{figures/layout_anno.png}
%     \caption{Snapshot the layout of \textit{Transporter}. At the center is the 128$\times$4 SPAD array. On the two sides of SPAD array are the passive quenching passive recharging circuits. }
%     \label{fig:enter-label}
% \end{figure}

\section{Simulation Results}

The chip will still be in the manufacturing process when this paper is published and will be tested in the near future. In the following subsections, the post-layout simulation results of the chip are shown, as well as the simulation of the SNN on the PC.

\subsection{Clock Module}

In the post-layout simulation, the ring oscillator generates a clock at 1.068 GHz (ss: 0.77 GHz, ff: 1.346 GHz). The simulation result of the gated clock is shown in Fig.~\ref{fig:sim-gclk}. One can observe the clock is disabled when the pulse count reaches 128 in every laser period (100 ns).

\subsection{SPAD and Pixel Circuits}

Given the operation voltage $V_{OP} = 25 \mathrm{V}$, breakdown voltage $V_B = 20 \mathrm{V}$, cascode voltage $V_{cascode} = 3.3 \mathrm{V}$, the quenching voltage $V_q$ is swept from 100 mV to 1 V with 10 equal steps. The result is shown in Fig.~\ref{fig:sim-pixel}. By adjusting $V_q$, the dead time can be tuned from a few nanoseconds to hundreds of nanoseconds.

\subsection{Transporter Ring}

The simulation result of the test structure for \textit{Transporter} ring is shown here. The collection of photons takes 2 $\mu$s, given a ring clock of 1 GHz. The readout of the spikes takes 1 $\mu$s, given a readout clock of 200 MHz. The laser period is 128 ns. With a simulated photon arrival with a period of 130 ns, the ring is supposed to receive the photons every two stages. The result is shown in Fig.\ref{fig:sim-readout}. During the accumulation period, 13 spikes are inserted into the ring, which are read out through the tri-state buffer bus. One can observe that the relative timing information is well preserved within the ring.

\subsection{Spiking Neural Networks for FLIM}

Based on the specifications of \textit{Transporter}, a spiking neural network (SNN) is designed, trained, and evaluated for fluorescence lifetime estimation. The SNN consists of a single-node input layer, a 512-neuron hidden layer, and a single-node output layer. Training is performed using backpropagation through time (BPTT) with surrogate gradients.

\begin{figure}
\centering
\subfloat[No background noise ]{
    \centering
    \includegraphics[width=0.23\textwidth]{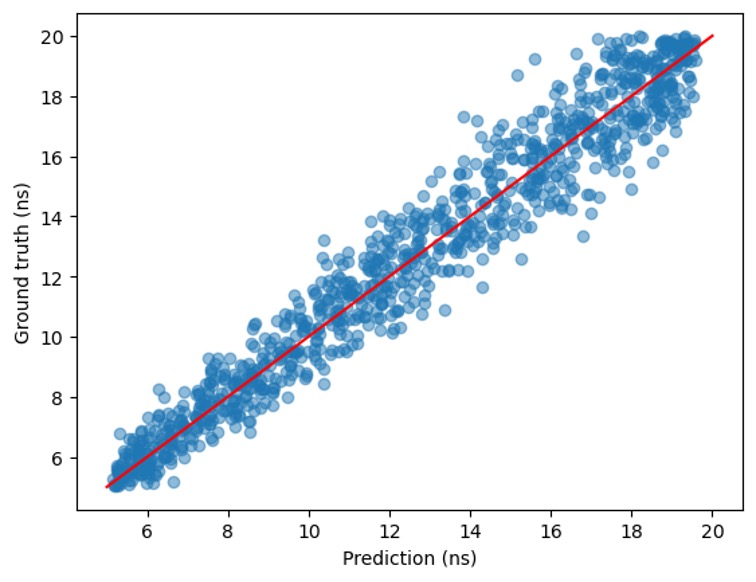}}
    \hfill
\subfloat[10\% background noise]{
    \centering
    \includegraphics[width=0.23\textwidth]{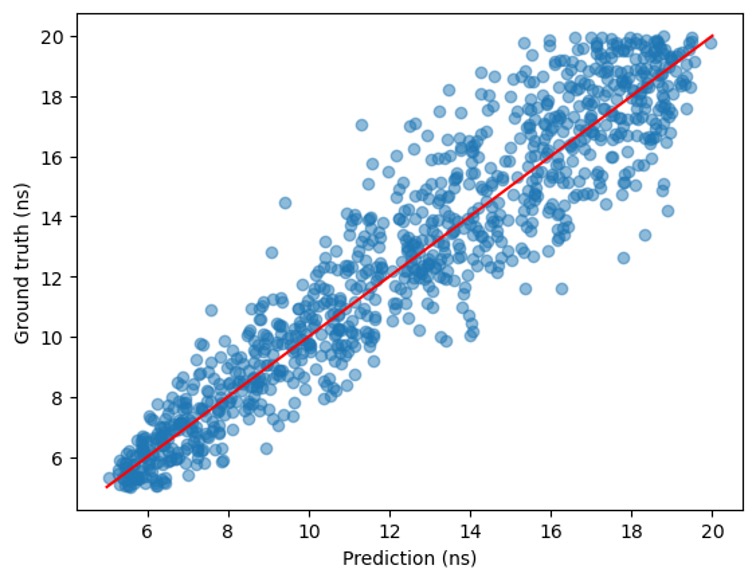}}
    \caption{Simulation results of spiking neural networks for fluorescence lifetime estimation. The parameters correspond to the ones in the \textit{Transporter}, i.e., timestep of 128 and clock period of 1 ns. The lifetime ranges from 5 to 20 ns. 256 photons are collected for each inference.}
    \label{fig:result}
\end{figure}

For the training and evaluation dataset, it is assumed that lifetime ranges from 5 to 20 ns. For each data entry, 256 timestamps are histogrammed and binarized as input for the SNN. Scenarios with no background noise and 10\% noise are considered. The result is shown in Fig.~\ref{fig:result}. The mean average percentage errors are 4.82\% and 5.29\%, respectively.

\section{Conclusion and Outlook}

In this work, we demonstrate the proof-of-concept \textit{Transporter} sensor that integrates a spike encoder for active time-resolved SPAD imaging, which explores a potential approach for realizing neuromorphic SPAD imaging systems. The suitability of \textit{Transporter} can be further enhanced by increasing the clock frequency, extending the encoder ring with more units, and utilizing simpler devices such as dynamic random access memory (DRAM). Our vision is to develop an extended encoder ring operating at a higher clock frequency, enabling real-time, TDC-free, and energy-efficient processing for advanced applications.

\section*{Acknowledgment}

The authors acknowledge the financial support provided by US Department of Energy.

% \section*{References}

{\small
\bibliographystyle{unsrt}
\bibliography{mybib}

@inproceedings{lin2024spiking,
  title={Spiking Neural Networks for Active Time-Resolved {SPAD} Imaging},
  author={Lin, Yang and Charbon, Edoardo},
  booktitle={Proceedings of the IEEE/CVF Winter Conference on Applications of Computer Vision},
  pages={8147--8156},
  year={2024}
}

@article{bruschini2019single,
  title={Single-photon avalanche diode imagers in biophotonics: review and outlook},
  author={Bruschini, Claudio and Homulle, Harald and Antolovic, Ivan Michel and Burri, Samuel and Charbon, Edoardo},
  journal={Light: Science \& Applications},
  volume={8},
  number={1},
  pages={87},
  year={2019},
  publisher={Nature Publishing Group UK London}
}

@article{kapusta2015advanced,
  title={Advanced photon counting},
  author={Kapusta, Peter and Wahl, Michael and Erdmann, Rainer},
  journal={Springer Series on Fluorescence},
  volume={15},
  year={2015},
  publisher={Springer}
}

@article{eshraghian2023training,
  title={Training spiking neural networks using lessons from deep learning},
  author={Eshraghian, Jason K and Ward, Max and Neftci, Emre O and Wang, Xinxin and Lenz, Gregor and Dwivedi, Girish and Bennamoun, Mohammed and Jeong, Doo Seok and Lu, Wei D},
  journal={Proceedings of the IEEE},
  volume={111},
  number={9},
  pages={1016--1054},
  year={2023},
  publisher={IEEE}
}

@inproceedings{milanesehistogram,
  title={Histogram-less direct time-of-flight imaging based on a machine learning processor on {FPGA}},
  booktitle={Proceedings of the International Image Sensor Workshop (IISW)},
  year={2023},
  author={Milanese, Tommaso and Zhao, Jiuxuan and Hearn, Brent and Charbon, Edoardo}
}

@article{lin2024coupling,
  title={Coupling a recurrent neural network to {SPAD} {TCSPC} systems for real-time fluorescence lifetime imaging},
  author={Lin, Yang and Mos, Paul and Ardelean, Andrei and Bruschini, Claudio and Charbon, Edoardo},
  journal={Scientific Reports},
  volume={14},
  number={1},
  pages={3286},
  year={2024},
  publisher={Nature Publishing Group UK London}
}
}

\afterpage{%
\definecolor{neoncarrot}{rgb}{1.0, 0.64, 0.26}
\tikzset{>=latex}
\usetikzlibrary{decorations.pathmorphing}

\definecolor{skyblue}{rgb}{0.53, 0.81, 0.92}

\definecolor{oceanboatblue}{rgb}{0.0, 0.47, 0.75}
\definecolor{napiergreen}{rgb}{0.16, 0.5, 0.0}
\definecolor{neoncarrot}{rgb}{1.0, 0.64, 0.26}

\tikzstyle{mynode}=[thick,draw=blue,fill=blue!20,circle,minimum size=22]

\tikzset{>=latex} % for LaTeX arrow head
\colorlet{myred}{red!80!black}
\colorlet{myblue}{blue!80!black}
\colorlet{mygreen}{green!60!black}
\colorlet{mydarkblue}{blue!40!black}
\tikzstyle{node}=[thick,circle,draw=myblue,minimum size=22,inner sep=0.5,outer sep=0.6]
\tikzstyle{node in}=[node,green!20!black,draw=mygreen!30!black,fill=mygreen!25]
\tikzstyle{node hidden}=[node,blue!20!black,draw=myblue!30!black,fill=myblue!20]
\tikzstyle{node out}=[node,red!20!black,draw=myred!30!black,fill=myred!20]
\tikzstyle{connect}=[thick,mydarkblue] %,line cap=round
\tikzset{ % node styles, numbered for easy mapping with \nstyle
  node 1/.style={node in},
  node 2/.style={node hidden},
  node 3/.style={node out},
}
\def\nstyle{int(\lay<\Nnodlen?min(2,\lay):3)} % map layer number onto 1, 2, or 3

\begin{figure*}[!tp]
    \centering
    \begin{circuitikz}[x=1cm,y=1cm,scale = 0.85, transform shape]
        % FFs
        \foreach [count=\ii] \N in {0,1,2,3,4,5,6,7}{
            \node[latch, scale=0.3, fill=skyblue](FF0-\ii) at (1.5*\N, 0) {};
            \node[latch, scale=0.3, rotate=180, rotated numbers, fill=skyblue](FF1-\ii) at (1.5*\N, -2) {};
        }
        % FF connections
        \foreach [count=\ii] \N in {0,1,2,3,4,5,6}{
            \pgfmathsetmacro\next{int(\ii+1)} % number of previous layer
            \draw[] (FF0-\ii.pin 6) -- (FF0-\next.pin 1);
            \draw[] (FF1-\next.pin 6) -- (FF1-\ii.pin 1);
        }

        \draw[] (FF0-8.pin 6) -| (11.7, -1) node[circ](readout){} |- (FF1-8.pin 1);

        \node[or port, scale=0.5, rotate=90, fill=skyblue](orgate) at (-1.5, -0.5) {};
        \node[flipflop SR, scale=0.5, fill=skyblue](ffsr) at (-2.5, -2) {};

        \draw[] (FF1-1.pin 6) -| (orgate.in 2) {};
        \draw[] (ffsr.pin 6) -| (orgate.in 1) {};
        \draw{} (orgate.out) |- (FF0-1.pin 1) {};

        % PLL

        \foreach [count=\ii] \N in {0,1,2,3,4,5,6,7}{
            \node[circ](conn-\ii) at (1.5*\N-0.75, -1) {};
            \draw[] (conn-\ii) |- (FF0-\ii.pin 3);
            \draw[] (1.5*\N+0.75, -1) |- (FF1-\ii.pin 3) {};
            \draw[] (1.5*\N-0.75, -1) -- (1.5*\N+0.75, -1) {};
        }

        \node[twoportshape, t=PLL, fill=skyblue](pll) at (1, -3.85) {};
        \draw[] (pll.w) -- (-0.75, -3.85) node[circ](pllo){} node[midway,above]{$Nf_0$};
        \draw[] (pllo) to[crossing] (conn-1);
        \node[right =1cm of pll](laser){Laser Sync};
        \draw (laser) -- (pll.e) node[above,midway]{$f_0$};
        \draw (pllo) -- (-3.5, -3.85) |- (ffsr.pin 3);

        % SPAD
        \ctikzset{ieeestd ports/pin length=0.5}
        \ctikzset{empty diodes}

        % \draw (0,3) to [pDo, fill=skyblue] (0, 4.5);
        \node[emptypdiodeshape, scale=0.7, rotate=90, fill=skyblue](pd) at (0, 3.7) {};
        \node[ieeestd not port, scale=0.5, fill=skyblue, rotate=180] (notgate) at (-1,3) {};
        \node[nmos, scale=0.7, rotate=180] (mosfet) at (0, 2.4) {};
        \node[right=0.1 cm of mosfet.G, scale=0.7] {$V_{bias}$};
        \draw (0,3) -- (mosfet.S);
        \draw (0,3) -- (notgate.west);
        \draw (0,3) -- (pd.west);
        \node[ground, scale=0.8](gnd) at (0,2) {};
        \node[vcc, above=0.5cm of pd.east, scale=0.7](vop){$V_{OP}$};
        \draw (pd.east) -- (vop);

        \draw [->, decorate, decoration={snake}, neoncarrot, thick] (-1.5, 4.8) -- (-0.7, 4.1);

        \draw (notgate) -- (-3.5, 3) |- (ffsr.pin 1);

        % NN
        \message{^^JNeural network with arrows}
        \readlist\Nnod{1,4,8,1} % array of number of nodes per layer
        
        \message{^^J  Layer}
        \foreachitem \N \in \Nnod{ % loop over layers
          \edef\lay{\Ncnt} % alias of index of current layer
          \message{\lay,}
          \pgfmathsetmacro\prev{int(\Ncnt-1)} % number of previous layer
          \foreach \i [evaluate={\y=\N/2-\i; \x=\lay; \n=\nstyle;}] in {1,...,\N}{ % loop over nodes
          
          % NODES
          \node[node \n,minimum size=11] (N\lay-\i) at (3+\x,3.5+0.5*\y) {};
          %\node[circle,inner sep=2] (N\lay-\i') at (\x-0.15,\y) {}; % shifted node
          %\draw[node] (N\lay-\i) circle (\R);
          
          % CONNECTIONS
          \ifnum\lay>1 % connect to previous layer
            \foreach \j in {1,...,\Nnod[\prev]}{ % loop over nodes in previous layer
              \draw[connect] (N\prev-\j) -- (N\lay-\i); % connect arrows directly
              %\draw[connect arrow] (N\prev-\j) -- (N\lay-\i'); % connect arrows to shifted node
            }
          \fi % else: nothing to connect first layer
          
        }
        }

        \draw (readout) -- ++(0.5,0) |- (N4-1);

        \draw[-{Stealth[length=2mm]}] (N1-1) -- ++ (-1.5, 0) -- ++(0, 2.15) node[above]{output};

        % boundary box
        \draw[dashed, line join=round] (-2, 5.4) -- ++(4, 0) node[midway,above=0.3cm] {SPAD} -- ++(0, -4.2) -- ++(-4, 0) -- ++(0, 4.2);

        \draw[dashed, line join=round] (3.5, 5.4) -- ++(4, 0) node[midway,above=0.3cm] {Spiking Neural Network} -- ++(0, -4.2) -- ++(-4, 0) -- ++(0, 4.2);

        \draw[line join=round] (8, 6) -- ++(4, 0) node[midway,above] {} -- ++(0, -2.5) -- ++(-4, 0) -- ++(0, 2.5);
        % \draw  (8, 6) -- (N3-1.north);
        % \draw  (8, 3.5) -- (N3-1.south);
        \draw (8, 4.75) -- (N3-1);

        \draw[dashed, line join=round] (-4, 0.9) -- ++(16, 0) -- ++(0, -5.5) -- ++(-16, 0)  node[midway,below] {Transporter Encoder} -- ++(0, 5.5);

        \begin{scope}[scale=0.5, shift={(17.5cm, 10cm)}, transform shape]
    \draw (4.5, 0) node[op amp, scale=0.5, fill=skyblue](comparator){};

    \draw (0, 52 |- comparator.+)node[mixer, scale=0.5, fill=skyblue](mymixer){};
    \draw [->] ($(-0.5,0.5)+(mymixer)$)node[left](){$w_i$} -- ++(0.25,0) -- (mymixer);
    \draw [->] ($(-0.5,-0.5)+(mymixer)$)node[left](){$t_{ij}$} -- ++(0.25,0) -- (mymixer);
    
    \draw (1.25, 52 |- comparator.+)node[adder, scale=0.5, fill=skyblue](myadder){};
    \draw (2.5, 52 |- comparator.+)node[adder, scale=0.5, fill=skyblue](myadder2){};

    % memory
    \ctikzset{multipoles/dipchip/width=3.5}
    \draw (2.5,-1.5) node[dipchip, num pins=10, hide numbers, no topmark, external pins width=0, scale=0.3](C){};
    \node [right, font=\tiny] at (C.bpin 3) {OUT};
    \node [left, font=\tiny] at (C.bpin 7) {RST};
    \node [left, font=\tiny] at (C.bpin 9) {IN};

    \node [below, text width=1.4cm, font=\small] at (C.south) {Potential Memory};

    \draw (comparator.out) -- ++ (0.5, 0) |- (C.bpin 7);
    \draw [->] (C.bpin 3) -| (myadder);
    \draw [->] (mymixer) -- (myadder);
    \draw [->] (myadder) -- (myadder2);
    \draw (myadder2) -- (comparator.+);

    \draw (3.6, 52 |- comparator.+) |- (C.bpin 9);
    \draw [->] (2.5, 1.2) node[above]{$\tau_{decay}$} -- (myadder2);
    \draw (3.6, 1.2) node[above]{$V_{thre}$} |- (comparator.-);

    \draw (myadder2.north) node[right] {-};
        \end{scope}
        
    \end{circuitikz}
    \caption{The proposed DFF ring-based \textit{Transporter} system. A SPAD with PQPR is adopted for photon detection. The pulse is injected into the ring, composed of $N$ DFFs (only 16 DFFs are plotted for simplicity), with an OR gate and keeps circulating. Ideally, the synchronization signal from the laser is divided by $N$ by a phase-locked loop (PLL) and drives all the DFFs. After the exposure, the clock is replaced by a low-speed readout clock and the pulses are read from the DFFs in serial.}
    \label{fig:transporter}
\end{figure*}
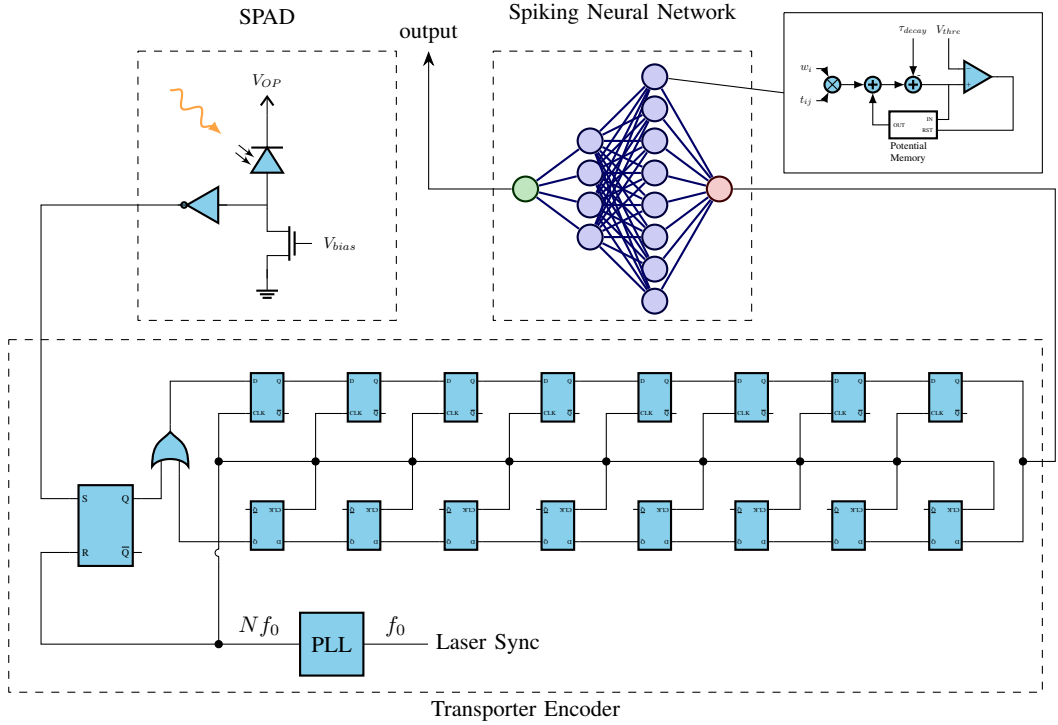
\begin{figure*}[!tp]
    \centering
    \subfloat[The output of the clock module with the internal ring oscillator and gating.]{
        \centering
    \includegraphics[width=\linewidth]{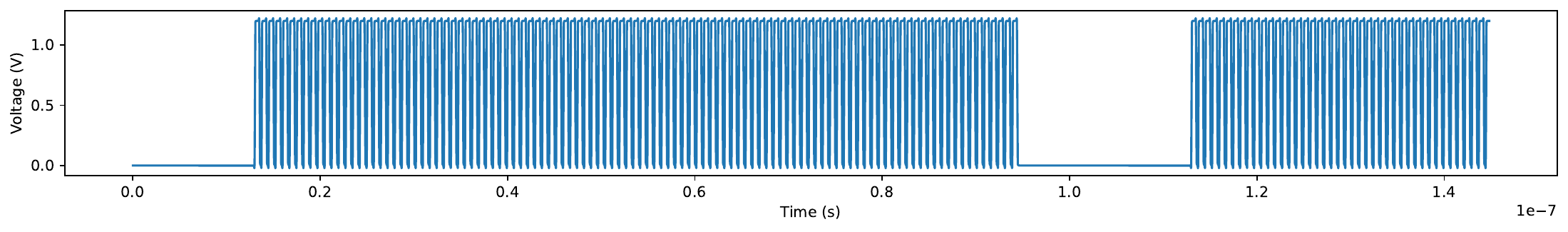}
    \label{fig:sim-gclk}
    }
    \\
    \subfloat[The anode and output of a SPAD with different $V_q$. The solid lines represent the voltage at the anode of the SPAD, and the dash lines represent the voltage digitized by inverters. Different colors represent different $V_q$.]{
        \centering
    \includegraphics[width=\linewidth]{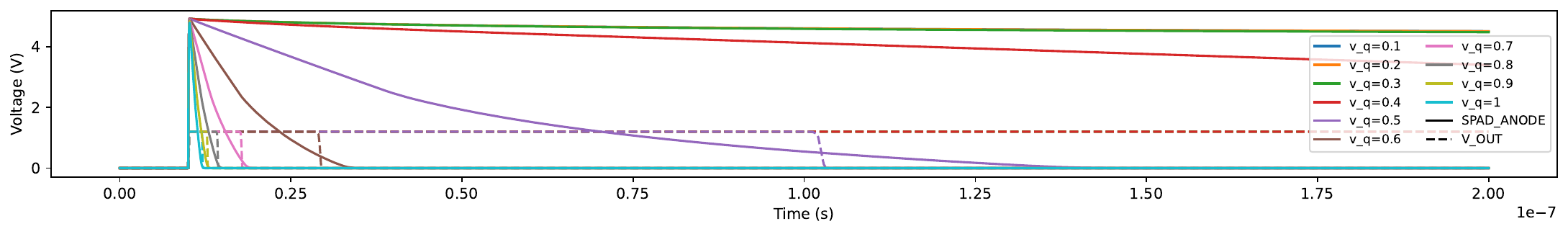}
    \label{fig:sim-pixel}
    }
    \\
    \subfloat[Readout of the spikes in the ring. The readout clock is 200MHz, so it takes 6.4 $\mu$s to read out 128 DFFs.  There are 13 spikes injected into the ring every two delay elements.]{
        \centering
    \includegraphics[width=\linewidth]{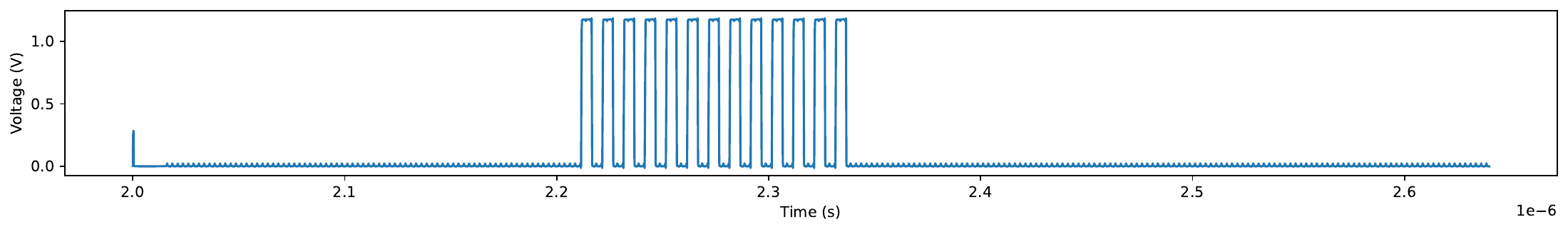}
    \label{fig:sim-readout}
    }
    \caption{Post-layout simulation results}
    % \label{fig:enter-label}
\end{figure*}
\clearpage
}

\end{document}